\documentclass[12pt]{article}

\usepackage{scicite}
\usepackage[superscript]{cite}

\usepackage{times}

\usepackage{graphicx}
\usepackage[displaymath]{lineno}

\newenvironment{sciabstract}{%
\begin{quote} \bf}
{\end{quote}}

\usepackage[usenames,dvipsnames]{xcolor}

\newcommand{\Ed}[1]{{\color{black}{#1}}} 

\newcounter{lastnote}

\title{\Ed{Bubble streams in Titan's seas as product of liquid N$_2$-CH$_4$-C$_2$H$_6$ cryogenic mixture}}

\author
{Daniel Cordier,$^{1\ast}$ Fernando Garc\'{i}a-S\'{a}nchez,$^{2}$\\ Daimler N. Justo-Garc\'{i}a,$^{3}$ G\'{e}rard Liger-Belair$^{1}$\\
\\
\normalsize{$^{1}$Groupe de Spectrom\'{e}trie Mol\'{e}culaire et Atmosph\'{e}rique - UMR CNRS 7331}\\
              \normalsize{Campus Moulin de la Housse - BP 1039,}\\
               \normalsize{Universit\'e de Reims Champagne-Ardenne}\\
               \normalsize{51687 REIMS -- France.}\\
\normalsize{$^{2}$Engineering Management of Additional Recovery,}
                  \normalsize{Mexican Petroleum Institute.}\\
                  \normalsize{Eje Central L\'{a}zaro C\'{a}rdenas Norte 152, 07730 Mexico City, Mexico.}\\
\normalsize{$^{3}$Department of Chemical and Petroleum Engineering,}\\
\normalsize{ESIQIE, National Polytechnic Institute.}\\
\normalsize{Unidad Profesional Adolfo L\'{o}pez Mateos,}\\
\normalsize{07738 Mexico City, M\'{e}xico}\\
\\
\normalsize{$^\ast$To whom correspondence should be addressed; E-mail:  daniel.cordier@univ-reims.fr.}
}

\date{}

\usepackage{color,xcolor}

\newcounter{firstbib}

\begin{document} 


\baselineskip24pt


\maketitle


\begin{sciabstract}
%
%
  Titan, Saturn's largest moon, is the only extraterrestrial body known to support stable
liquid on its surface, in the form of \Ed{seas and lakes that dot the polar regions}. 
Many indications suggest that the liquid should be composed of a mixture of
N$_2$, CH$_4$ and C$_2$H$_6$. 
Recent RADAR observations of \Ed{Titan's large sea, called ``{\it Ligeia Mare}'',} have shown unexplained and
ephemeral bright features, possibly due to rising bubbles. 
\Ed{Here we report that our numerical model, when combined with experimental data found in the literature, 
shows that Ligeia Mare's bed is a favourable place for nitrogen exsolution.}
This process could produce centimeter-sized and radar-detectable bubbles.
\end{sciabstract}


%
%
\section*{Introduction}

     Titan, the main satellite of the giant planet Saturn, is the only moon in the solar
system harboring a dense atmosphere. Among many unique features, this gaseous envelope, mainly composed by nitrogen and methane,
is the place of a ``hydrological'' cycle of methane. In addition, the complex photochemistry of its atmosphere produces 
a wide variety of organic species with ethane being, in terms of quantity, its main product{\cite{yung_etal_1984,krasnopolsky_2009}}.
%
%
The RAdio Detection And Ranging (RADAR) instrument, onboard the Cassini spacecraft, allowed for the discovery of dark features, partially covering Titan's polar 
regions. Long suspected to be there (refs{\cite{flasar_1983,lunine_etal_1983}}), they are interpreted as seas or lakes of 
liquid hydrocarbons {\cite{stofan_etal_2007}}. While their exact chemical composition
is not known, the lower atmosphere contains around $5$\% of CH$_4$ and $95$\% of N$_2$ and
together with C$_2$H$_6$, produced by photochemistry, these species should be the main components of Titan's seas 
{\cite{cordier_etal_2009,tan_etal_2013,brown_etal_2008}}. Titan's maritime environments exhibit an absence of ocean waves 
{\cite{zebker_etal_2014}}, but
strange transient RADAR over-brightness events have been reported in two places at the surface of the northern sea, Ligeia Mare. 
In order to explain these episodes, the formation of gas bubbles has been proposed together with
potential suspended/floating solids{\cite{hofgartner_etal_2014,hofgartner_etal_2016}}. In this work, we focus on thermodynamic 
instabilities that can produce a nitrogen exsolution, and consequently feed streams of bubbles, which can explain the events observed by RADAR.\\
%

%
\section*{Existence of Liquid-Liquid-Vapor Equilibria for hydrocarbon mixtures}

   The most straightforward inter-phases equilibrium, of a N$_2$-CH$_4$-C$_2$H$_6$ ternary system, is a liquid-vapor equilibrium
(LVE). However, liquid-liquid-vapor equilibria (LLVE) do occur under certain conditions in ternary
and liquefied natural gas (LNG) systems, including liquid phase inversion {\cite{ramirez-jimenez_etal_2012}}.
LLVE equilibria consist in the coexistence of two liquids, of different compositions, with
a vapor. 
In the laboratory, LLVEs have been observed 
for systems, under cryogenic conditions, comparable to Titan's liquid phases: nitrogen + methane + (ethane, or propane, or  $n$-butane) 
{\cite{ramirez-jimenez_etal_2012,lu_etal_1970,yu_1972,merrill_etal_1984,llave_etal_1987}}. 
%
%
%

     At the {\it Huygens} landsite, {\it i.e.} in Titan's tropical regions, the temperature was $93.65 \pm 0.25$ K {\cite{fulchignoni_etal_2005}}. 
Using the CIRS (The Composite Infrared Spectrometer) instrument aboard Cassini, Titan's atmospheric temperature has been 
retrieved as a function of latitude, for the lowest $1$ km. The average temperature of the northern polar region, during
mid-northern spring, has been found around $91$ K\cite{jennings_etal_2016}{ }. However, the question of the specific temperature at the surface 
of Ligeia Mare remains
open, since the thermal properties of the liquid could differ, compared to those of surrounding lands. Moreover, the behavior of temperature with 
depth is not well constrained. Most estimations rely only on numerical simulations that depend on many parameters, like the light extinction
coefficient of the sea {\cite{legall_etal_2016,tokano_lorenz_2016}}. Nevertheless, all models agree with a with a temperature cooling of a few degrees between
the surface and the sea bottom {\cite{tokano_2009,tokano_lorenz_2016}}. We therefore adopted a characteristic range of Ligeia Mare temperatures
from $80$ to $90$ K.\\
%
     Titan's ground pressure, measured {\it in situ} by the {\it Huygens} probe, is close to $P_{0}=1.5$ bar and should not undergo significant variations
over the satellite surface {\cite{fulchignoni_etal_2005}}. 
Since the maximum depths of Ligeia Mare have been determined to be between $\sim 100$ m and 
$\sim 200$ m{\cite{mastrogiuseppe_etal_2014,hayes_2016,legall_etal_2016}}, we estimated the pressure at these depths.
  As a first approximation, liquids can be considered as incompressible. In that frame, the pressure $P$ at a given depth $z$ within a Titan's sea follows
the law
\begin{equation}
  P= P_{0} + \bar{\rho} \, g_{\rm Tit} \, z
\end{equation}
where $\bar{\rho}$ represents the mean density of the liquid between the considered depth and the surface, $g_{\rm Tit}$ denotes the gravity.
Values for the density of liquid CH$_4$, C$_2$H$_6$ or N$_2$ can be found in the literature{\cite{younglove_ely_1987,handbook74th}}, for instance under
$1$ bar $\rho_{\rm CH_4}= 451.8$ kg.m$^{-3}$, $\rho_{\rm C_2H_6}= 651.6$ kg.m$^{-3}$, while $\rho_{\rm N_2}\sim 800$ kg.m$^{-3}$. We check that these
values are only weakly dependent of pressure: \textit{e.g.} at $10$ bars $\rho_{\rm C_2H_6}= 651.8$ kg.m$^{-3}$ and $\rho_{\rm CH_4}= 452.1$ kg.m$^{-3}$.
Using these numbers we found $P_{100} \simeq 2.1$ bar at $100$ m (\textit{resp}. $P_{200}= 2.7$ bar at $200$ m) for pure methane layers. For layers composed exclusively
of ethane the pressure at $100$ m reaches $P_{100}= 2.4$ bar (\textit{resp.} $P_{200}= 3.3$ bar). Due to the higher density of N$_2$ ($870$ kg.m$^{-3}$ at 
its triple point where $T_{t}= 63$ K, probably $\sim 800$ kg.m$^{-3}$ at $\sim 90$ K), some amount of nitrogen would increase the mean density of the mixture, 
and then, the pressure at $100$ or $200$ m. If we considere the heaviest hydrocarbon \Ed{under consideration}, \textit{i.e.} ethane, and if we consider a nitrogen mean abundances of $40$\% 
(in mole fraction), which represents
probably a very high solubility of N$_2$, we found $P_{100}= 2.5$ bar and $P_{200}= 3.5$ bar. In addition to these estimations, we developed a more sophisticated
model based on the law of hydrostatics and the equation of state (EoS) PC-SAFT {\cite{gross_sadowski_2001,tan_etal_2013}}(see Methods). This way, the compressibility
of the fluid was taken into account. We also investigated the influence of temperature in the range $80-90$ K. With this model, we arrived to the same conclusion: 
the pressure at $100-200$ m, should be in the range $2-3.5$ bar.

%
%
%
%
\section*{Stability Analysis of the Ternary System CH$_4$-N$_2$-C$_2$H$_6$}

   The most relevant laboratory experiments of LLVEs have been acquired at $94.3$ K under a pressure between $4.2$ and $4.55$ bars {\cite{yu_1972}}. 
These \Ed{conditions are} inconsistent with Ligeia Mare's expected bathymetry and temperature profile.
  As a consequence, we investigated the stability of the N$_2$-CH$_4$-C$_2$H$_6$ system at a lower pressure, and cooler temperature. To do so, we used
a numerical method allowing phase stability analysis, originally introduced for the industry of gas and oil. 
%

   First, we have investigated whether the lake's surface could be the location of phase splitting events, \textit{i.e.} the appearence of two liquids of
distinct compositions, in equilibrium with the atmosphere. For a temperature
and pressure appropriate for Titan's surface, \textit{i.e.} $90$ K and $1.5$ bars, we found that --by varying the mole fraction
of nitrogen-- the binary system N$_2$+CH$_4$ does not undergo any demixing, it remains either in vapor or in a vapor-liquid equilibrium. 
Same results were obtained for N$_2$+C$_2$H$_6$. For the ternary mixture N$_2$+CH$_4$+C$_2$H$_6$, 
we have followed two distinct scenarios, in order to maximise the range of parameter space we explore: (1) we fixed the abundance in N$_2$, 
of the entire system (\textit{i.e.} including the vapor and the liquid(s)) and only varied the abundances of CH$_4$ from $0.05$ to $0.001$ 
\Ed{(letting ethane making up the remainder)}
in mole fraction and (2) we fixed the overall composition of CH$_4$ and gradually changed the fraction of N$_2$ from $0.95$ to $0.90$. 
All cases resulted in an simple liquid-vapor equilibrium. The vapor, representing the
Titan's atmosphere, was always largely dominated by N$_2$.
In the light of these results, we can safely conclude that, at the expected polar ground conditions, the Titan's seas surface is thermodynamically stable 
and should not split into two liquids.\\
%

   Next, we searched for LLVE for conditions relevant to environments below the sea surface (pressure up to $\sim 3.5$ bars and 
temperature possibly down to $\sim 80-85$ K). Our results are summarized in Fig. 1. For all three considered temperatures, we found LLVE.
In all cases, these three-phases equilibria consist in two liquids in coexistence with a vapor phase composed almost exclusively of nitrogen (empty squares 
in Fig. 1). The existence of this almost pure nitrogen vapor requires the exsolution of some amount of N$_2$ contained in the liquid before the occurrence
of the LLVE. One liquid is a nitrogen-rich phase (filled squares in Fig. 1); the other is enriched in ethane (empty circles in Fig. 1). 
We have also evaluated the densities of the
phases involved in our study. A set of values, corresponding to a specific LLVE, is presented in the insert included in Fig. 1. The higher density of the 
nitrogen-rich liquid is a trend confirmed for all the LLVEs we found.
Low temperature and high pressure both favor the occurrence of a phase splitting that results in a LLVE. At $T= 90$ K, 
the temperature at the surface of the sea or slightly below, 
a LLVE appears at a pressure as low as $\sim 2.7$ bars. These conditions allow the occurrences of LLVEs at depths between $130-170$ m, 
compatible with Ligeia Mare's bathymetry. Clearly, the lower temperature, due --for instance-- 
to infrared absorption in layers near the surface, would favor phase splitting at much shallower depths in the range $20$--$30$ m (see Fig. 1 panels a
and b). The temperature strongly influences the depth required to obtain a pressure high enough to reach conditions for an exsolution.\\
%
%
%
\section*{Implication for Titan}

     A sea with a homogeneous composition that matches the one required for the occurrence of a LLVE, at some depth, is an improbable scenario.
In addition, such a case would imply a nitrogen degassing through the whole extent of the system. Instead, we propose a scenario where the sea is
vertically stratified with ethane enriched bottom layers and methane-rich upper layers, the latter containing more
dissolved nitrogen. This is supported by the higher density of ethane-rich mixtures and the larger solubility of nitrogen in methane than in ethane 
{\cite{tan_etal_2013}}. In addition, surface layers are in contact with the nitrogen rich atmosphere, situation that favors the dissolution of N$_2$. 
In addition, given Ligeia Mare's surface are of several thousand square kilometers, local episodes of evaporation and precipitation of methane 
can lead to a horizontal gradient of compound abundances
{\cite{tokano_lorenz_2016}}.
3D modeling of ocean circulation with a modified version of the Bergen Ocean Model suggests
the existence of maritime streams produced by tides, wind or
generated by solar heating{\cite{tokano_etal_2014,tokano_lorenz_2016}}. We propose that the vertical sea circulation 
feeds locally the lowest sea layers with methane and nitrogen enriched
liquid. Depending on how exactly the composition has been altered at the surface (by local rain, evaporation or N$_2$ dissolution), the downward flow 
can meet the deep ethane-rich layers, this fullfilling the composition, pressure and temperature required for the occurrence of a LLVE.
Our thermodynamically modeling informs us
only about which equilibria are possible under given conditions, but is unable to predict either the involved 
amounts of matter, or the kinetics of the described process. Nevertheless, in our scenario, circumstances favorable to LLVE, are ephemeral: composition 
of liquid sinking from the surface has a composition varying with time, first due to weather over the sea, and second due to mixing during downward 
flow. Once the phase splitting begins, at
a specific reached depth, the droplets of the denser liquid (nitrogen-rich) tend to sink while the lighter ones (ethane-rich) rise toward the surface.
Of course, this phases separation process tends to stop naturally the formation of a LLVE, by the exhaustion of one or several required constituents.
Both kind of mentioned droplets should redissolve into the ambient liquid during their descent or ascent, because the pressure and temperature move away 
from those
required for a LLVE. It is worth noting that buyoancy-driven bubbles of almost pure nitrogen rapidly rise through the upper N$_2$-rich layers
to the surface, and for this reason probably do not redissolve into surrounding liquid. This scenario matches the Ligeia Mare observations 
of RADAR over-brightness events.
It is also striking that the locations of the Transient Feature Ligeia 1 (TFL1) and TFL2{\cite{hofgartner_etal_2016}}, 
superimposed to the depth map indicate places at the 
boundaries of sea areas deeper than $\sim 100$ m. Moreover, it should be noted 
that bubbles are probably not observed at the surface, right above of their \Ed{formation site},
due to their own stochastic movement and to sea circulation. The break-up diameter of buoyancy-driven gas bubbles, which has to be understood as a 
maximum diameter, can be estimated as
\begin{equation}
   d_{\rm breakup} \sim 8 \sqrt{\frac{\gamma}{g_{\rm Tit} \Delta\rho}}
\end{equation}
 \Ed{where} $\gamma$ is the surface tension of the liquid, $g_{\rm Tit}$ stands for the Titan's surface gravity and 
 $\Delta\rho= \rho_{liq}-\rho_{gas}$ with $\rho_{liq}$
and $\rho_{gas}$, respectively, the density of the surrounding liquid and the gas held by \Ed{the} bubbles {\cite{clift_etal_1978}}. 
\Ed{Using} surface tensions \Ed{from} the Dortmund Data Bank \Ed{(\texttt{http://www.ddbst.com})}, and for a mixture
of CH$_4$, C$_2$H$_6$ and N$_2$ (0.4:0.4:0.2), composition that could be typical of the upper layers of liquid, at $90$ K 
$\gamma_{\rm mix} \sim 2 \times 10^{-2}$ N m$^{-1}$; $\Delta\rho$ is estimated at a few bars for the same mixture 
thanks to PC-SAFT. This yields a $d_{\rm breakup} \sim 4.6$ cm, approximately twice the RADAR wavelength ($2.2$ cm). It should be noted that a composition 
based only on CH$_4$ and N$_2$ with a ratio $1:1$, which is probably an extreme situation, leads also to a $d_{\rm breakup}$ of a few centimeters.
Such large bubbles can magnify the Cassini RADAR backscattering, allowing easy detection, similarly to what is suspected for pebbles in dry river beds on 
Titan \cite{legall_etal_2010} or evaporites {\cite{cordier_etal_2016b}}. In comparison, predicted sea current speeds, of a few cm s$^{-1}$, can lift only 
sediment particles well below $1$ cm in size {\cite{tokano_lorenz_2016}}, and, then cannot be detectable with the Cassini RADAR.\\
%

    At its specific site of occurrence, the LLVE demixing should perturb the local circulation of fluids. Indeed, the densest liquid sinks below the lightest 
while vapor rises to the surfaces. This phase separation tends also to stop the process, since it changes locally the composition. Unfortunately, our 
approach can not provide the amount of matter involved in the different phases. No quantification can be then proposed concerning the influence on 
circulation and the magnitude of degassing. 
Of course, we have focused on Ligeia because ``Magic Islands'' (\textit{i.e.} nicknames for TFL1 and TFL2) were observed at this sea. 
But, our argumentation can be applied to other mare like Kraken and Punga. However, the existence of depths between $100$ and $200$ m is a 
crucial point. We can reasonably think that such depths exist under the surface of Kraken, but in the case of Punga it is much more questionable
as the bathymetry is not yet available and as Punga could be substantially less deep than Ligeia and Kraken, as it is suggested by its smaller size.
%
%
Since, nitrogen exsolution in the deepest parts of Titan's seas can represent a potential hazard for an in situ exploration {\cite{hartwig_etal_2016,lorenz_etal_2015}}, 
future laboratory experiments, specifically designed to study these ternary equilibria, 
would be extremely useful. As a priority, the existence of LLVEs, found in past experiments has to be confirmed by modern measurements{\cite{hollyday_etal_2016,farnsworth_etal_2017}}. These
investigations could then be easily extended to the pressures of a few bars higher than the Titan's surface one, and to temperature down to a few degrees
below $\sim90$ K. This way, our numerical stability analysis could be reinforced. Concerning the seas circulation, 3D models including a full
treatment of the chemical composition, with the possibility of LLVEs, would be welcome. But, since data used as inputs of such models, and in the first
place the sea beds topography, are still poorly known, the last word will come from a Titan's submarine exploration.\\
%
%
%
%

\newpage

\noindent\textbf{\Large Author contributions}\\
D. Cordier wrote the paper and performed PC-SAFT computations, F. Garc\'{i}a-S\'{a}nchez and D. N. Justo-Garc\'{i}a
made the stability analysis of the N$_2$-CH$_4$-C$_2$H$_6$ mixtures, G. Liger-Belair provides expertise concerning the physics of bubbles and 
effervescence.\\


\noindent\textbf{\Large Additional information}\\
\noindent \textbf{Supplementary information} is available for this paper.\\
\noindent \textbf{Reprints and permissions information} is available at \texttt{www.nature.com/reprints}.\\
\noindent \textbf{Correspondence and requests for materials} should be addressed to D.C.\\


\noindent\textbf{\Large Competing interests}\\
\noindent The authors declare no competing financial interests.\\

\clearpage

\noindent {\bf Fig. 1.} The behavior of a ternary mixture N$_2$-CH$_4$-C$_2$H$_6$ at three temperature values relevant for Titan's 
          sea subsurface environment: (a) $T= 80$ K, (b) $T= 85$ K and (c) $T= 90$ K. Each set of circles, filled squares and empty squares represents
          a LLVE for a given pressure. 
          While the LLVE shown at $T= 80$ K is an example that illustrates the influence of temperature, panels (b) and (c) provide estimations 
          of the lowest pressures at which LLVE occur at $T= 85$ K and $T= 90$ K, respectively $\sim 1.7$ bar and $\sim 2.7$ bar.         
          A grey-shaded panel provides composition and densities of 
          fluids involved in the LLVE corresponding to $T= 85$ K and $P= 1.8$ bar (see panel b).
%
\begin{figure}[!t]
\begin{center}
\includegraphics[width=10 cm]{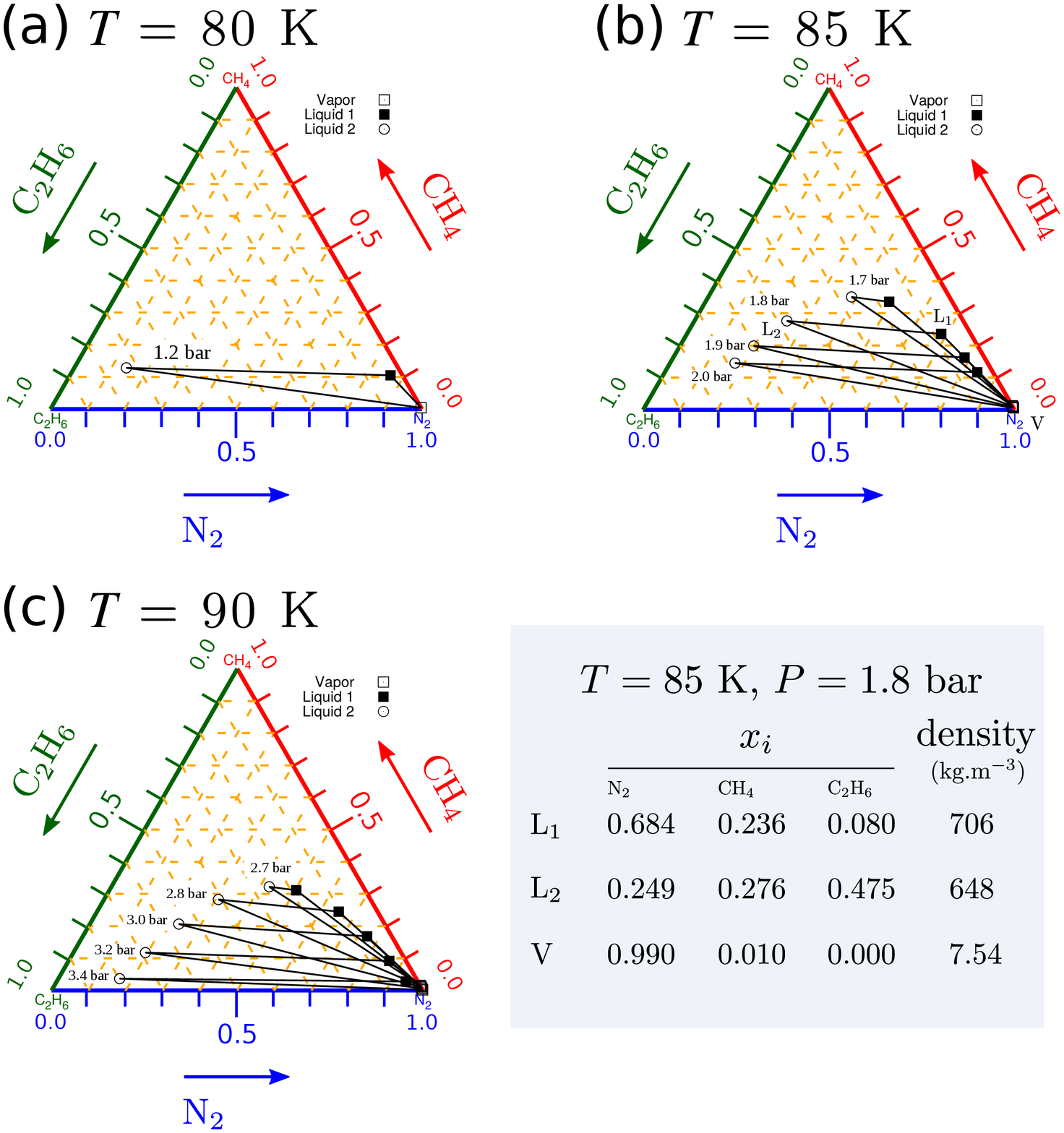}
\end{center}
\end{figure}
%
%
%
\newpage
\section*{Methods}

   In order to evaluate the pressure at the bottom of Titan's seas we used the well known equation
\begin{equation}
\frac{\displaystyle\partial P}{\displaystyle\partial z}= -\rho g_{\rm Tit}
\end{equation}
  where $z$ is the depth, $\rho$ the density and $g_{\rm Tit}$ the satellite ground gravity. The density depends on the pressure $P$, the temperature $T$
and the mole fractions of the considered species N$_2$, CH$_4$ and C$_2$H$_6$. The PC-SAFT equation of state (EoS) {\cite{gross_sadowski_2001}}, widely employed in 
the chemical engineering community, was used to determine the density $\rho$. 
In the PC-SAFT EoS, the molecules are conceived to be chains composed of spherical segments, in which the pair potential for the segment of 
a chain is given by a modified square-well potential {\cite{chen_kreglewski_1977}}. Non-associating molecules are characterized by three pure-component parameters: 
the temperature independent segment diameter $\sigma$, the lenght of the potential $\epsilon/k_{\rm B}$, and the number of segments per chain $m$. The PC-SAFT 
EoS, expressed in terms of the Helmholtz energy for a multicomponent mixture of non-associating chains, consists of a hard-chain reference contribution and a 
perturbation contribution to account for the attractive interactions. 
The three pure-component parameters of nitrogen, methane and ethane for the PC-SAFT EoS were taken from the literature {\cite{gross_sadowski_2001}}. 
For mixtures, the PC-SAFT EoS uses classical van der Waals one-fluid mixing rules for the perturbation terms. In these mixing rules, the parameters for a pair 
of unlike segments are obtained trough conventional Lorentz-Betherlot combining rules, where one binary interaction parameter $k_{ij}$ is introduced to correct 
the segment-segment interactions of unlike chains. 
The binary interaction parameters used in all the phase equilibrium calculations for the PC-SAFT EoS are: 
$0.0307$ for N$_2$-CH$_4$, $0.0458$ for CH$_4$-C$_2$H$_6$, and $-0.0058$ for CH$_4$-C$_2$H$_6$, and they were also taken from the literature 
{\cite{garcia-sanchez_etal_2004,justo-garcia_etal_2008}}.\\
   Concerning the stability analysis, the employed technique relies on the above mentioned PC-SAFT EoS,
and uses an efficient computational procedure for solving the isothermal multiphase problem.
Initially, the system is assumed to be monophasic. A stability test allows checking whether the system is stable or not. If stability is reached, the procedure is stopped.
In the contrary case, it provides an estimation of the composition of an additional phase to take into account for the equilibrium calculation. The number of phases is 
then increased by one, and equilibrium is reached by minimizing the Gibbs energy. The procedure is continued until a stable solution is found {\cite{ramirez-jimenez_etal_2012}}. 
%
   This approach has been validated by comparison with many laboratory measurements involving ternary mixtures. For instance, a very good agreement is found 
with mixtures of liquid methane, nitrogen and $n$-pentane, or $n$-hexane or $n$-heptane at various temperatures {\cite{justo-garcia_etal_2009}}. In addition, in the specific 
context of this work, we have payed great attention to the reproduction of laboratory data, for the system N$_2$-CH$_4$-C$_2$H$_6$ at $94.3$ K under $\sim 4$ bars, 
reported in Ref.~{\cite{yu_1972}}.\\

\noindent\textbf{\Large Data availability.} The data that support the plots within this paper and
other findings of this study are available from the corresponding author upon
reasonable request.\\

\end{document}